\documentclass[aps,prb,reprint]{revtex4-2}
\usepackage{amsmath}
\usepackage{graphicx}
\usepackage{units}  
\usepackage{xspace}
\usepackage{subfigure}
\usepackage{hyperref}

\newcommand{\ve}{\varepsilon}
\newcommand{\mb}{\mathbf}

\newcommand{\tb}{\textbf}
\newcommand{\beq}{\begin{equation}}
\newcommand{\eeq}{\end{equation}}
\newcommand{\bea}{\begin{eqnarray}}
\newcommand{\eea}{\end{eqnarray}}

\usepackage[usenames]{color}

\begin{document}

\bibliographystyle{apsrev}
 
\title{Quantum Contribution to Magnetotransport in Weak Magnetic Fields and Negative Longitudinal Magnetoresistance}

\date{\today}

\author{Hridis K. Pal}
   \email{hridis.pal@iitb.ac.in}
\affiliation{Department of Physics, Indian Institute of Technology Bombay, Powai, Mumbai 400076, India}

\begin{abstract}
Longitudinal magnetoresistance (LMR) refers to the change in resistance due to a magnetic field when the current and the magnetic field are parallel to each other. For this to be nonzero in weak magnetic fields, kinetic theory stipulates that the electronic dispersion must satisfy certain conditions: it should either be sufficiently anisotropic or have topological features. The former results in a positive LMR while the latter results in a negative LMR. Here, I propose a different mechanism that leads to LMR in \emph{any} dispersion without a need to satisfy the above requirements. The mechanism is quantum in origin but is applicable in the said regime. It arises due to the change in the density of states with the magnetic field and is not kinetic in origin. Remarkably, LMR is found to be negative even if the dispersion is nontopological, provided it is nonparabolic. An analytical expression is derived for this novel contribution to LMR. It  is found to depend on the orbital magnetic susceptibility. The analytical findings are confirmed by numerical calculations. 
\end{abstract}

\maketitle 

\section{Introduction}

Magnetotransport---the motion of charge carriers in the presence of both an electric and a magnetic field---is one of the most commonly studied phenomena in solids. As a charge accelerates under the electric field, it suffers repeated collisions with scatterers, giving rise to resistance. With the introduction of the magnetic field, the charge now experiences an additional Lorentz force and bends  away from its linear path. It spends more time in traversing the direction of the electric field and suffers more collisions resulting in an increased resistance. Thus, magnetoresistance is expected to be positive and arise only when the current and the magnetic field have components perpendicular to each other \cite{zim}. 

Based on the above kinetic picture which is essentially classical, one does not expect longitudinal magnetoresistance (LMR) to exist because it requires the current to be parallel to the magnetic field; and even if it exists, it should not be negative (N). Nevertheless,  in solids where electrons do not have a free-particle dispersion, LMR---and in some cases NLMR---can arise within the same basic mechanism at weak fields where kinetic theory is valid. It has been shown that if the dispersion possesses a certain kind of anisotropy such that the velocity of electrons parallel and perpendicular to the magnetic field can not be decoupled, LMR can be nonzero and is necessarily positive \cite{pal,pipbook,pip}. On the other hand, if the dispersion features topological properties, along with the Lorentz force the kinetics is influenced by an additional contribution that arises from the Berry curvature. Then anisotropy is not a necessity and a nonzero LMR can arise. In this case, however, LMR is negative \cite{nielsen,burkov,son,niu,andreev}. Apart from these two mechanisms, others that lead to LMR---and in some cases NLMR---are known to exist. However, these are either extrinsic in origin, such as scenarios that require specific models of scattering \cite{sondheimer,goswami} and inhomogeneities \cite{stroud,miller,parish}, or are beyond the semiclassical regime requiring, for example, a very high magnetic field which forces electrons to occupy only the lowest Landau level \cite{argyres} or low enough temperatures such that quantum interference effects lead to weak (anti)localization \cite{abrbook}.

In this work I show that there exists another mechanism, intrinsic in origin and applicable in weak fields but not kinetic in nature, that contributes to magnetotransport. This gives rise to a nonzero value of LMR  in cases where kinetic theory predicts a zero value, which can even become negative. The mechanism derives from the change of density of states due to the magnetic field and is quantum in origin in spite of appearing in a classical regime. A simple understanding can be obtained by considering a familiar context in which the same mechanism is at play: Landau diamagnetism. It is well known that classically Landau diamagnetism cannot arise since the magnetic field through its kinetic contribution can not affect the total energy of a system. Quantum mechanically, however, it is allowed since the density of states becomes a function of the magnetic field through the formation of discrete Landau levels. Note that, in spite of being quantum in origin, the effect manifests at weak fields such that $\omega_c\ll E_F$, where $\omega_c$ is the cyclotron frequency and $E_F$ is the Fermi energy ($\hbar=1$). Extending this mechanism to transport, magnetoresistance should also inherit a similar contribution, irrespective of the orientation of the current and the magnetic field. Importantly, the orbital magnetic susceptibility, while being diamagnetic for a parabolic dispersion (Landau diamagnetism), becomes paramagnetic when nonparabolicity is introduced in the dispersion \cite{fred}. The same can arise in the context of magnetotransport with magnetoresistance switching sign from positive to negative as the dispersion acquires nonparabolicity. The different physical origins of the two contributions to magnetoresistance, kinetic and quantum, are expected to show up in their functional dependence on the magnetic field: the former is expected to be a function of $\omega_c\tau$, where $\tau$ is the relevant scattering time, whereas the latter should be a function of $\frac{\omega_c}{E_F}$. Because $\omega_c\tau$ can reach values much larger than one while satisfying $\omega_c\ll E_F$, in general the kinetic contribution will dominate over the quantum contribution. However, if the former is identically zero, the latter can become the leading contribution. As discussed earlier, this could happen with LMR making the the new quantum contribution relevant in this context. 

Below I substantiate the above claims with analytical and numerical calculations. LMR is calculated for a dispersion that is separable in directions parallel and perpendicular to the magnetic field using the Kubo formula. The choice of such a dispersion is not necessary, but is done for two reasons: first, it greatly simplifies the calculation, and second, it is known that the kinetic contribution to LMR for such a choice is identically zero \cite{pal}; therefore, any LMR found is necessarily of quantum origin. A general expression for the quantum contribution is derived which is found to be intimately related to the orbital magnetic susceptibility. It is explicitly shown that, in contrast to conventional wisdom, even a parabolic dispersion exhibits LMR, which becomes negative as nonparabolicity is introduced in the dispersion.

\section{Model}

\begin{figure}
\includegraphics[angle=0,width=0.8\columnwidth]{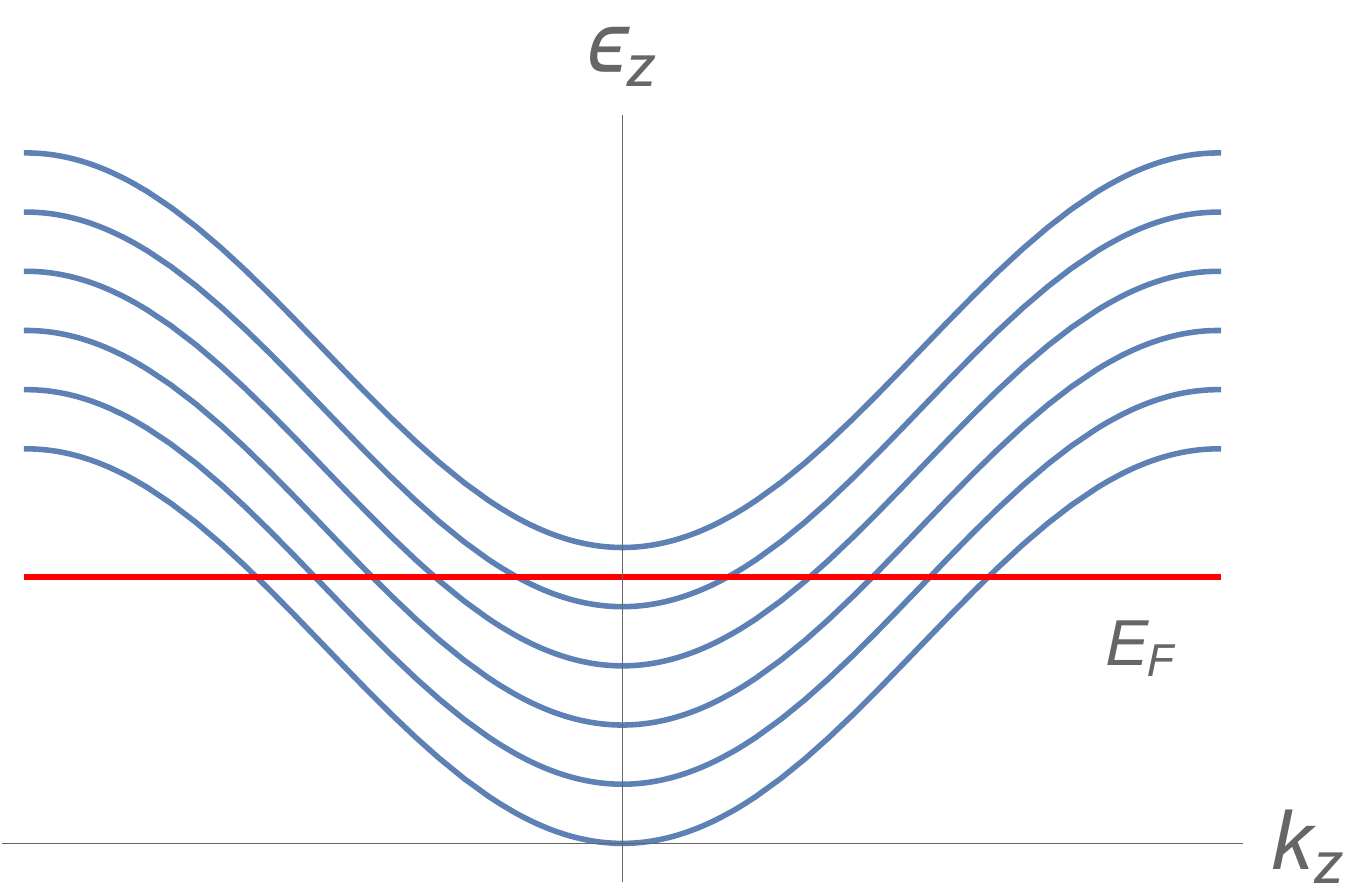}
\caption{In a magnetic field, a three-dimensional spectrum is reduced to a set of one-dimensional bands dispersing in the direction of the field. The sum in Eq.~(\ref{sigzz}) is over those bands that intersect the Fermi energy $E_F$.}
 \label{figband}
\end{figure}

Consider a metallic system with a dispersion
\beq
E(\mb{k})=\ve_{xy}(k_x,k_y)+\ve_{z}(k_z).\label{dispk}
\eeq
Without any loss of generality it is assumed that the minimum value of each term is zero. A magnetic field $\mb{B}$, described by the vector potential $\mb{A}=(0,Bx,0)$, is applied in the $z-$direction. The dispersion becomes (spin is ignored for simplicity)
\beq
E(n,k_z)=\ve_{xy}(n)+\ve_{z}(k_z),\label{dispersion}
\eeq
where $\ve_{xy}(n)$ denotes the Landau levels in two dimensions. The eigenfunctions are given by
\beq
\psi_{k_z,k_y,n}=e^{ik_zz+ik_yy}\phi_n(x-k_yl_B^2),\label{eigenfunc}
\eeq
where $\phi_n$ are the Landau levels eigenfunctions corresponding to $\ve_{xy}(n)$ and $l_B^2=\frac{1}{eB}$. The corresponding single particle Green's function is
\beq
G(k_z,k_y,x,x',\omega)=\sum_{n}\frac{\phi_n^\ast(x-k_yl_B^2)\phi_n(x'-k_yl_B^2)}{\omega-\xi_n(k_z)+\frac{i}{2\tau}sgn(\omega)},\label{green}
\eeq
where $\xi_n(k_z)=E(n,k_z)-E_F$. Here, I have included a phenomenological scattering time $\tau$ without worrying about its microscopic origin and assumed it to be field independent. This will be revisited later. Throughout this work, it will be assumed that scattering is weak so that $1/\tau\rightarrow 0$.

\section{Longitudinal magnetoconductivity}

In calculating the longitudinal magnetoconductivity $\sigma_{zz}$, I closely follow Abrikosov \cite{abr} who first calculated the same for a parabolic dispersion. Using the Kubo formula, 
\begin{widetext}
\beq
\sigma_{zz}(B)=\mathrm{Re}\frac{e^2}{(2\pi)^3}\int d\omega\frac{n_F(\omega)-n_F(\omega+\Omega)}{\Omega}\int dk_zdk_ydx'
v_{z}(k_z)G^R(k_z,k_y,x,x',\omega)v_{z}(k_z)G^A(k_z,k_y,x',x,\omega+\Omega).\label{kubo}
\eeq
\end{widetext}
Here, $n_F$ is the Fermi function, $v_z=\frac{\partial E}{\partial k_z}=\frac{\partial \ve_z}{\partial k_z}$, $G^{R(A)}$ is the retarded (advanced) Green's function corresponding to Eq.~(\ref{green}), and $\Omega$ is the external frequency. At $T=0, \Omega=0$, the frequency integral pins all energies on the Fermi surface. Using Eq.~(\ref{green}) in  Eq.~(\ref{kubo}), I have
\begin{eqnarray}
&&\sigma_{zz}(B)
=\frac{e^2}{(2\pi)^3}\sum_{n,n'}\int dk_zdk_ydx'\frac{ v_z^2(k_z)}{\xi^2_n(k_z)+\frac{1}{4\tau^2}}\nonumber\\
&&\phi_n^\ast(x-k_yl_B^2)\phi_n(x'-k_yl_B^2)\phi_{n'}^\ast(x'-k_yl_B^2)\phi_{n'}(x-k_yl_B^2).\nonumber
\end{eqnarray}
Using the fact that the Landau level eigenfunctions form an orthonormal complete basis, the integral over $x'$ gives $\delta_{n,n'}$. Using this, and completing the integral over $k_y$, I have
\beq
\sigma_{zz}(B)=\frac{e^2eB}{(2\pi)^3}\sum_{n}\int dk_z \frac{v_z^2(k_z)}{\xi^2_n(k_z)+\frac{1}{4\tau^2}}.
\eeq
Next, I make a change of variable: $\int dk_z\rightarrow\int\frac{2}{|v_z|}d\xi_n$, where the factor 2 is included since $E$ is an even function of $k_z$ (see comment \cite{comment}). In the limit $\frac{1}{\tau}\rightarrow 0$, $\frac{1}{\xi^2_n(k_z)+\frac{1}{4\tau^2}}\rightarrow 2\tau\pi\delta(\xi_n)$. Thus,
\beq
\sigma_{zz}(B)=\frac{e^2\tau eB}{2\pi^2}\sum_{n}\lvert v_{zn}\rvert,\label{sigzz}
\eeq
where 
\beq
v_{zn}=\left.\frac{\partial\ve_{z}(k_z)}{\partial k_z}\right\rvert_{k_z=\ve_{z}^{-1}(E_F-\ve_{xy}(n))\ge 0}.\label{vz}
\eeq
The summation over $n$ runs from $0$ to $N$, the maximum value of $n$ for which $\ve_{xy}(n)\le E_F$. Equation~(\ref{sigzz}) has a simple interpretation. The magnetic field has reduced the three-dimensional spectrum into a set of  one-dimensional bands dispersing along $k_z$, each with a degeneracy proportional to $B$. The total conductivity is the sum of the velocity in the $z-$direction at the Fermi energy contributed by all the partially occupied bands. As shown in Fig.~\ref{figband}, the number of such bands is given simply by the number of bands $E_F$ crosses---this corresponds to $N$. At very high magnetic fields, only the lowest band is occupied ($N=1$) which contributes to transport. This is a purely quantum regime. As the magnetic field decreases, more bands get populated by going below the Fermi level. When the number of occupied levels is large ($N\gg 1$) one enters the semiclassical regime. In this regime, with change in $B$, the sum in Eq.~(\ref{sigzz}) changes in two ways: a part that evolves smoothly and another that changes abruptly due to a sudden change from $N$ to $N+1$ each time an extra band gets populated. Together, they give rise to LMR, the former appearing as a smooth background while the latter manifesting as quantum oscillations. Quantum oscillations are vestigial signatures of quantum effects in the semiclassical regime. The fully quantum regime along with quantum oscillations in the semiclassical regime have been extensively studied before by Arbrikosov \cite{abr} and others \cite{argyres,sho} as manifestations of quantum effects. However, the smooth background is considered to be purely classical, described by the kinetic theory, devoid of any quantum effects. Below, through explicit calculations I show that this is not correct: the smooth background contribution to LMR also inherits an intrinsic quantum contribution, hitherto unexplored, with novel consequences.

As a simple example consider first a parabolic spectrum: $\ve_{xy}(k_x,k_y)=\frac{k_x^2+k_y^2}{2m}$ giving $\ve_{xy}(n)=(n+\frac{1}{2})\omega_c$, where $\omega_c=\frac{eB}{m}$, and $\ve_{z}(k_z)=\frac{k_z^2}{2m}$. Then, 
$v_{zn}=\sqrt{\frac{2}{m}}\sqrt{E_F-\left(n+\frac{1}{2}\right)\omega_c}.$
The summation over $n$ in Eq.~(\ref{sigzz}) can be converted into an integral by using the Euler-MacLaurin formula (see Supplemental Material). Ignoring the oscillating part and keeping only the smooth part upto $\mathcal{O}(B^2)$, I find 
\beq
\sigma_{zz}(B)\approx\sigma_{zz}(0)\left[1-\frac{1}{32}\frac{\omega_c^2}{E_F^2}\right],\label{parasigzz}
\eeq
where $\sigma_{zz}(0)=\frac{n_0e^2\tau}{m}$, $n_0=\frac{(2mE_F)^{3/2}}{6\pi^2}$ being the zero-field charge density. Thus, even for a parabolic spectrum the longitudinal conductivity is magnetic field dependent. This should be contrasted with the kinetic theory result which predicts absence of any field- dependence. The field-dependent part scales with $\frac{\omega_c}{E_F}$ instead of $\omega_c\tau$, confirming its quantum origin. The negative sign implies that LMR, obtained by taking the inverse, is positive.

\begin{figure}
\includegraphics[angle=0,width=0.95\columnwidth]{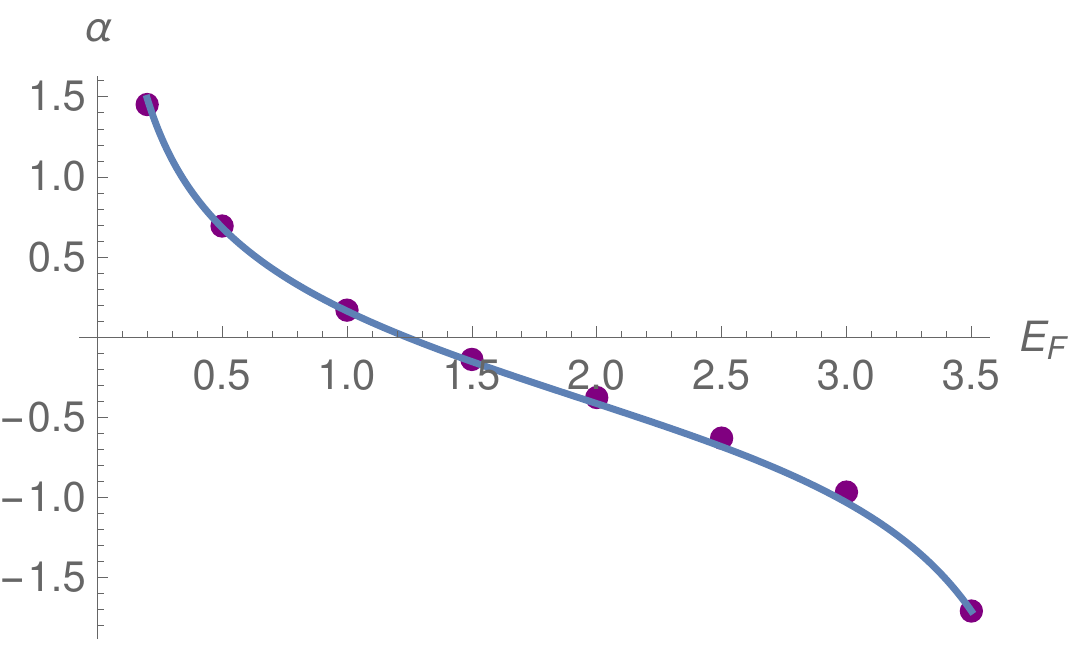}
\caption{Dependence of $\alpha$ (in arbitrary units) on $E_F$ (in units of $t$) in longitudinal conductivity, $\sigma_{zz}(B)=\sigma_{zz}(0)-\alpha B^2$ for the separable spectrum in Eq.~(\ref{dispk}) with $\ve_{xy}=4t-2t[\mathrm{cos}(k_xa)+\mathrm{cos}(k_ya)]$ and $\ve_{z}=\frac{k_z^2}{2m}$. The solid line is according to the analytical expression in Eq.~(\ref{result}). The solid circles represent numerically calculated values obtained by fitting the curves in Fig.~\ref{figLMR} to the quadratic equation above. At small $E_F$, the spectrum is close to parabolic and $\alpha$ is positive. However, as $E_F$ is increased, the spectrum becomes nonparabolic and $\alpha$ becomes negative---in this regime, LMR is negative.}
 \label{lmrsign}
\end{figure}

I now generalize the above idea to a general spectrum. The Landau levels $\ve_{xy}(n)$ no longer have a simple analytical form. They are, instead, derived from the semiclassical quantization condition,
\beq
Sl_B^2=2\pi(n+\gamma),
\eeq
where $S(\ve)$ is the area of the surface enclosed by the isoenergy contour $\ve_{xy}=\ve$ in the two-dimensional $k-$space and $\gamma$ is the semiclassical phase. It is easy to check that for a parabolic dispersion, $S(\ve)=\pi (k_x^2+k_y^2)|_{\ve_{xy}=\ve}=2\pi m\ve$ and $\gamma=\frac{1}{2}$ reproduce the correct Landau level spectrum $\ve(n)=(n+\frac{1}{2})\omega_c$. When the dispersion is non-parabolic, two changes arise: $S(\ve)$ is no longer the area of a circle and, more importantly, $\gamma$ is no longer a constant but a function of $\ve$ itself. While $S(\ve)$ is a simple geometrical quantity, calculation of $\gamma(\ve)$ requires more care. In the simplest case where singularities in the isoenergy contours and interband effects can be ignored, it was shown by Roth that \cite{roth,sho}
\beq
\gamma(\ve)-\frac{1}{2}=\frac{eB}{48\pi}\frac{\partial}{\partial\ve}\int \delta(\ve_{xy}-\ve)\left[m_{xx}^{-1}m_{yy}^{-1}-(m_{xy}^{-1})^2\right]d^2k,\label{roth}
\eeq
where $m_{\alpha\beta}^{-1}=\frac{\partial^2\ve_{xy}}{\partial k_\alpha\partial k_\beta}$. This can be written in terms of the two-dimensional orbital magnetic susceptibility $\chi$. According to the Landau-Peierl's formula \cite{roth},
\beq
\chi(\ve)=-\frac{e^2}{24\pi^2}\int \delta(\ve_{xy}-\ve)\left[m_{xx}^{-1}m_{yy}^{-1}-(m_{xy}^{-1})^2\right]d^2k.\label{lp}
\eeq
Combining the two, 
\beq
\gamma(\ve)-\frac{1}{2}=-\frac{\pi B}{2e}\frac{\partial \chi}{\partial\ve}.\label{gammafinal}
\eeq
Going back to Eq.~(\ref{sigzz}), the sum is once again computed using the Euler-Maclaurin formula, but keeping in mind that now a change in $n$ is accompanied by changes in both $S$ and $\gamma$ \cite{sho}. Ignoring the oscillating part and keeping only the smooth part upto $\mathcal{O}(B^2)$ as before, I find
\begin{eqnarray}
\sigma_{zz}(B)&\approx&\sigma_{zz}(0)-\left[\frac{\partial |v_z|}{\partial\ve}\chi\bigg\rvert_{\ve=0}+\int_{0}^{E_F}\frac{\partial |v_z|}{\partial\ve}\frac{\partial\chi}{\partial\ve}d\ve\right]\frac{e^2\tau}{4\pi} B^2\nonumber\\\label{result}
&=&\sigma_{zz}(0)-\alpha B^2.
\end{eqnarray}
Here, $\frac{\partial |v_z|}{\partial\ve}\equiv\frac{\partial |v_z|}{\partial\ve_{xy}}\rvert_{\ve_{xy}=\ve}$, where $v_z$ is evaluated from Eq.~(\ref{dispersion}) and expressed in terms of $\ve_{xy}$ [similar to Eq.~(\ref{vz}) but now in $(k_x,k_y)$ space].  The expression for LMR is obtained by inverting Eq.~(\ref{result}): $\rho_{zz}(B)\approx\rho_{zz}(0)+\alpha B^2$, where $\rho_{zz}=\frac{1}{\sigma_{zz}}$. Equation~(\ref{result}) clearly shows that the quantum contribution to LMR in a three-dimensional system is intimately related to the orbital magnetic susceptibility of the corresponding two-dimensional spectrum, confirming their common origin. 

A remarkable feature of Eq.~(\ref{result}) is that the two terms constituting the coefficient $\alpha$ need not be of the same sign; therefore, $\alpha$ can pick a sign depending on which term wins. In the parabolic case, $v_z=\frac{k_z}{m}=\sqrt{\frac{2}{m}}\sqrt{E_F-\ve}$ and $\chi=-\frac{e^2}{12\pi m}$ [from Eq.~(\ref{lp})]. The latter is independent of energy, so the second term constituting $\alpha$ drops out and the expression in Eq.~(\ref{parasigzz}) is recovered with $\alpha$ positive. However, once the dispersion becomes nonparabolic, the second term becomes nonzero and opposite in sign to the first term. For a sufficiently nonparabolic spectrum, $\alpha$ becomes negative resulting in NLMR. Note that for this to happen, it is sufficient to have only the two-dimensional spectrum $\ve_{xy}$ nonparabolic, the dispersion along the magnetic field, $\ve_{z}$ can still be parabolic. To illustrate this, consider the spectrum $\ve_{xy}=4t-2t[\mathrm{cos}(k_xa)+\mathrm{cos}(k_ya)]$ and $\ve_{z}=\frac{k_z^2}{2m}$, where $t$ is the nearest neighbor hopping parameter on a square lattice of lattice constant $a$. Using Eq.~(\ref{lp}), one finds \cite{fred} $\chi(\ve)=\frac{e^2}{12ta^2\pi^2}Q_{1/2}\left[1-\frac{(\ve-4)^2}{8}\right]$, where $Q_{n}[x]$ is the Legendre function of the second kind and $\ve$ is in units of $t$. Using this in Eq.~(\ref{result}), the integral is calculated to compute $\alpha$. In Fig. \ref{lmrsign}, the dependence of $\alpha$ on $E_F$ is plotted. At small $E_F$, the spectrum is close to parabolic, and $\alpha$ is positive. With increase in $E_F$, nonparabolicity becomes more pronounced and at some value $\alpha$ switches sign and becomes negative, resulting in NLMR. Equation (\ref{result}) along with its consequences form the main result of this paper.

\begin{figure}
\includegraphics[angle=0,width=0.95\columnwidth]{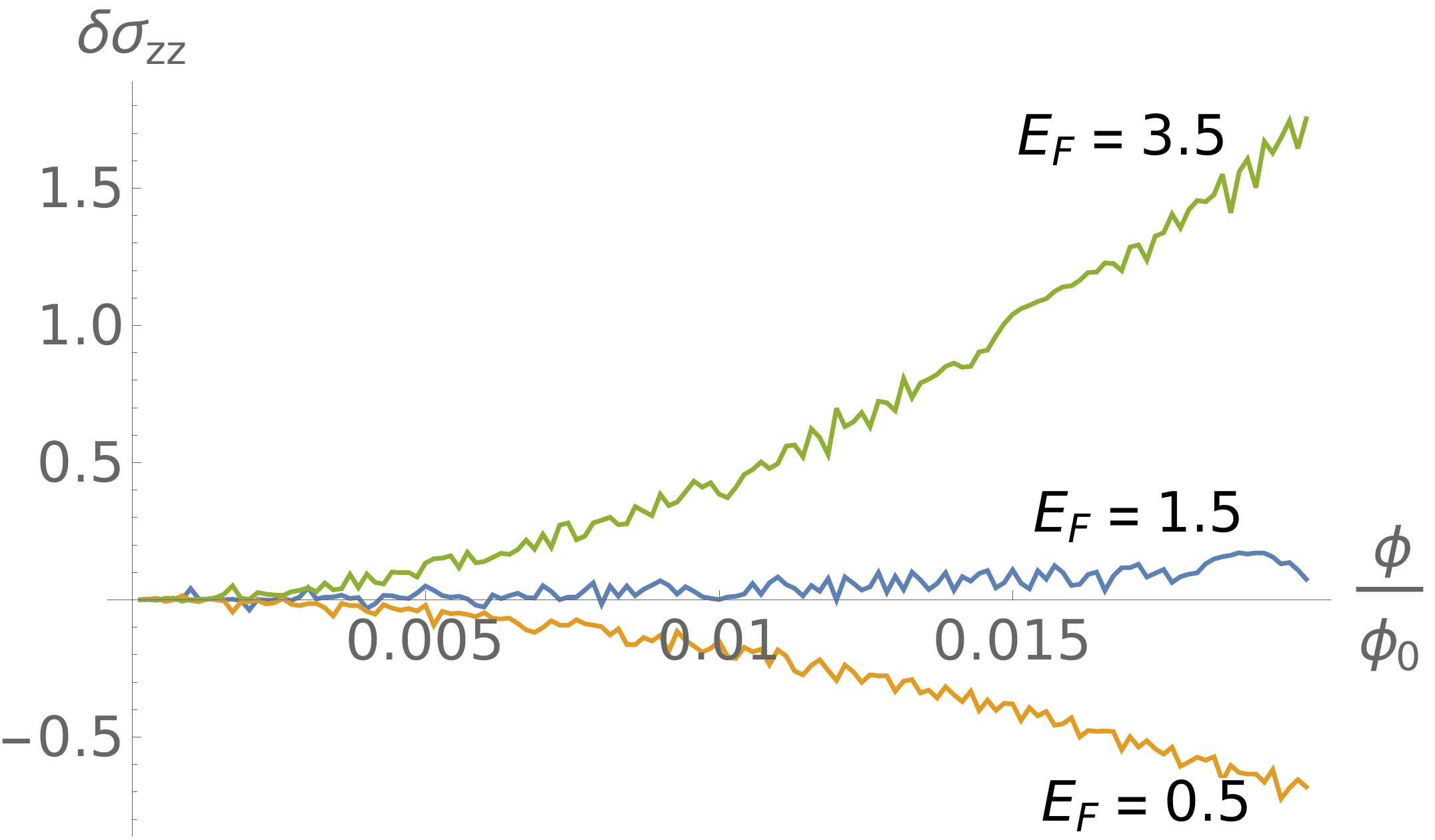}
\caption{Dependence of $\delta\sigma_{zz}=\sigma_{zz}(B)-\sigma_{zz}(0)$ (in arbitrary units) on $B$ (expressed in terms of flux over flux quantum) at different values of $E_F$ (in units of $t$) obtained by evaluating Eq.~(\ref{sigzz}) exactly---see text. The spectrum is same as in Fig.~\ref{lmrsign}: $\ve_{xy}=4t-2t[\mathrm{cos}(k_xa)+\mathrm{cos}(k_ya)]$ and $\ve_{z}=\frac{k_z^2}{2m}$. A nonzero temperature $T=0.1 t$ has been used to suppress the quantum oscillations. At small values of $E_F$, $\sigma_{zz}$ decreases with $B$ whereas at larger values of $E_F$, it increases---in this regime, LMR is negative.}
\label{figLMR}
\end{figure}

\section{Numerical calculation}

As further proof I now present an exact numerical evaluation of Eq.~(\ref{sigzz}), which is then compared with the analytical result in Eq.~(\ref{result}). The Landau level spectrum $\ve_{xy}(n)$ corresponding to $\ve_{xy}(k_x,k_y)=4t-2t[\mathrm{cos}(k_xa)+\mathrm{cos}(k_ya)]$ is calculated numerically on a lattice model (see Supplemental Material for details). Using Eq.~(\ref{vz}),  $v_{zn}=\sqrt{\frac{2}{m}}\sqrt{E_F-\ve_{xy}(n)}$ is computed. This is inserted in Eq.~(\ref{sigzz}) and the sum is evaluated numerically as a function of the field. This yields the total $\sigma_{zz}(B)$ which includes both the smooth as well as the oscillating parts. To remove the oscillating part, a small temperature is introduced. Temperature influences the two contributions differently: it introduces a negligible correction $\sim\left(\frac{T}{E_F}\right)^2$ (Sommerfeld correction) in the smooth part, but reduces the oscillating part exponentially as $\sim e^{-T/\omega_c}$ for $T\gg \omega_c$. This is exploited to suppress the oscillating part and reveal the smooth part of $\sigma_{zz}(B)$. Note that, this is not just a theoretical trick, but also has experimental relevance: to observe the predicted behavior in the smooth part of LMR, one needs to be in the regime $\omega_c\lesssim T\ll E_F$. The effect of temperature is included by using the formula $\sigma_{zz}(E_F,T)=\int\left(-\frac{\partial n_F(E-E_F)}{\partial E}\right)\sigma_{zz}(E,0)dE$. The results are presented in Fig. \ref{figLMR}. As expected, $\sigma_{zz}(B)$ varies quadratically with the field. At small values of $E_F$ it decreases with the field leading to positive LMR while at larger values of $E_F$ it becomes an increasing function of the field leading to NLMR. The curves are fitted and the coefficient $\alpha$ is extracted. The extracted values of $\alpha$ are plotted in Fig. \ref{lmrsign} alongside the analytical curve. It is seen that they are in excellent agreement.

\section{Effect of field on scattering time}

In arriving at Eq.~(\ref{sigzz}) the scattering time $\tau$ was assumed to be a phenomenological constant. In a microscopic theory, $\tau$ depends on the density of states and, therefore, should change with the field. More importantly, since $\tau$ is inversely proportional to the density of states, one can wonder whether it will kill all the field dependence in $\sigma_{zz}$ found so far. It turns out that this is not the case. This can be shown explicitly by considering a simple model where delta-function impurities are scattered randomly in a system with a parabolic spectrum. Assuming weak and dilute impurities, within the first Born approximation one finds (see Supplemental Material) $\tau^{-1}=n_iU_0^2\frac{eB}{\pi}\sum_{n=0}^N\frac{1}{|v_{zn}|}$, where $U_0$ is the Born scattering amplitude and $n_i$ is the density of impurities. Inserting this in Eq.~(\ref{sigzz}) it is clear that a cancellation does not occur. Carrying out the summation over the Landau levels as before (see Supplemental Material), I find $\tau\approx\tau_0\left[1-\frac{1}{96}\frac{\omega_c^2}{E_F^2}\right]$, where $\tau_0$ is the scattering time in the absence of the field. Using this in Eq.~(\ref{parasigzz}), I get $\sigma_{zz}(B)\approx\sigma_{zz}(0)\left[1-\frac{1}{24}\frac{\omega_c^2}{E_F^2}\right]$. The field dependence in $\tau$, instead of destroying LMR, accentuates it.

\section{Concluding remarks}

To summarize, I have shown that a nonzero LMR can arise in \emph{any} dispersion in weak magnetic fields, in contrast to the prediction of kinetic theory which states that LMR is nonzero only for dispersions of certain kinds. This arises because  a magnetic field affects electronic transport not only kinetically, but also by modifying the density of states. The mechanism is inherently quantum in spite of manifesting in the classically weak-field regime.  Importantly, the quantum contribution to LMR can become negative if the dispersion is sufficiently nonparabolic, even if the latter has no topological features. It is found that it is related to the orbital magnetic susceptibility. While the theory presented here considered the simplest case of a single isolated band, it can be extended to include coupled bands. Such extensions are important in the context of topological systems and will be investigated in future.

\begin{acknowledgments}
I am grateful to F. Pi{\'e}chon for valuable discussions. Part of this work was completed during my visit to LPS, Orsay, France in 2019 which was partially supported by the French program LabEx PALM Investissement d'avenir (ANR-10-LABX-0039-PALM) within the project TOPOMAGTRANS. I thank IRCC, IIT Bombay for financial support via grant RD/0518-IRCCSH0-029.
\end{acknowledgments}

\begin{widetext}

\section*{Supplemental Material}

\subsection{Calculation of $\sigma_{zz}$ for a parabolic spectrum}

\subsubsection{Constant $\tau$}

For the separable energy spectrum in a magnetic field,
\beq
E(n,k_z)=\ve_{xy}(n)+\ve_{z}(k_z),\label{dispersion_supp}
\eeq
the longitudinal magnetoconductivity is given by
\beq
\sigma_{zz}(B)=\frac{e^2\tau eB}{2\pi^2}\sum_{n}|v_{zn}|,\label{sigzz_supp}
\eeq
where 
\beq
v_{zn}=\left.\frac{\partial\ve_{z}(k_z)}{\partial k_z}\right\rvert_{k_z\rightarrow \ve_{z}^{-1}(E_F-\ve_{xy}(n))}.\label{vz_supp}
\eeq
I first assume the dispersion to be parabolic: $\ve_{xy}(n)=(n+1/2)\omega_c$ and $\ve_{z}(k_z)=k_z^2/2m$, with $\omega_c=eB/m$. This gives $v_{zn}=\sqrt{\frac{2}{m}}\sqrt{E_F-(n+1/2)\omega_c}$, and Eq.~(\ref{sigzz_supp}) becomes
\beq
\sigma_{zz}(B)=\frac{e^2\tau}{2\pi^2}\omega_c\sqrt{2m}\sum_{n=0}^N\sqrt{E_F-(n+1/2)\omega_c}.\label{sigzz_supp2}
\eeq
To calculate the discrete sum, I use the Euler-Maclaurin formula
\beq
\sum_{r=0}^Rf(r)=\int_0^Rf(r)dr+\frac{1}{2}[f(R)+f(0)]+\frac{1}{12}[f'(R)-f'(0)]+\cdots.
\eeq
For convenience, define $n+1/2=x$ and $X=E_F/\omega_c$, i.e., the value $x$ takes at $E_F$, and $\delta=X-(N+1/2)$. Then,
\begin{eqnarray}
\sum_{n=0}^N\sqrt{E_F-\left(n+\frac{1}{2}\right)\omega_c}&=&\int_{\frac{1}{2}}^{X-\delta}\sqrt{E_F-x\omega_c}dx+\frac{1}{2}\left[\sqrt{E_F-(X-\delta)\omega_c}+\sqrt{E_F-\frac{1}{2}\omega_c}\right]\nonumber\\
&+&\frac{1}{12}\left[(\sqrt{E_F-x\omega_c})'\rvert_{x\rightarrow X-\delta}-(\sqrt{E_F-x\omega_c})'\rvert_{x\rightarrow 1/2}\right]+\cdots\nonumber\\
&=&-\frac{2}{3\omega_c}\left[(\delta\omega_c)^{3/2}-(E_F-\omega_c/2)^{3/2}\right]+\frac{1}{2}\left[(\delta\omega_c)^{1/2}+(E_F-\omega_c/2)^{1/2}\right]\nonumber\\
&-&\frac{\omega_c}{24}\left[\frac{1}{(\delta\omega_c)^{1/2}}-\frac{1}{(E_F-\omega_c/2)^{1/2}}\right]+\cdots.
\end{eqnarray}
Expanding in $\omega_c/E_F$,
\beq
\sum_{n=0}^N\sqrt{E_F-\left(n+\frac{1}{2}\right)\omega_c}\approx\left[\frac{2}{3}\frac{E_F^{3/2}}{\omega_c}\right]+\left[-\frac{1}{48}\frac{\omega_c}{E_F^{1/2}}\right]+\left[\frac{\omega_c^{1/2}}{\delta^{1/2}}\left(-\frac{2}{3}\delta^{2}+\frac{1}{2}\delta-\frac{1}{24}\right)\right].
\eeq
Using this in Eq.~(\ref{sigzz_supp2}),  I finally have
\beq
\sigma_{zz}(B)\approx\frac{e^2\tau}{2\pi^2}\frac{2^{3/2}m^{1/2}}{3}E_F^{3/2}\left[1-\frac{1}{32}\frac{\omega_c^2}{E_F^2}-\frac{\omega_c^{3/2}}{E_F^{3/2}}\frac{1}{\delta^{1/2}}\left(\delta^2-\frac{3}{4}\delta+\frac{1}{16}\right)\right].\label{parabolares_supp}
\eeq
The first term is the zero-field contribution, the second is the smooth part that varies quadratically with the field, and the third leads to oscillations. I discard the oscillating part and take the first two terms, which appear in the main text.

\subsubsection{Field dependence of $\tau$}

In the previous calculation $\tau$ was considered to be some phenomenological constant. Here, I derive it in the simple case of weak and dilute disorder with delta-function impurities scattered randomly. The self-energy $\Sigma$ considering the simplest diagram within the first Born approximation (Fig.~\ref{selfen_supp}) is given by
\beq
\Sigma=n_{i}U_0^2\frac{1}{(2\pi)^2}\sum_{n=0}^{N}\int\frac{\phi_n^\ast(x-k_yl_B^2)\phi_n(x-k_yl_B^2)}{\omega-\xi_n(k_z)+i\eta\mathrm{sgn}(\omega)}dk_ydk_z,
\eeq
where $U_0$ is the Born scattering amplitude and $n_{i}$ is the density of impurities, and $\eta\rightarrow 0^{+}$. Completing the integral over $k_y$ yields a prefactor of $l_B^2=eB$. Using $\int dk_z\rightarrow \int\frac{2}{|v_z|} d\xi_n$, I have
\beq
\Sigma=-i\mathrm{sgn}(\omega)n_iU_0^2\frac{eB}{2\pi}\sum_{n=0}^{N}\frac{1}{|v_{zn}|}.\label{sigma}
\eeq
The self-energy turns out to be purely imaginary because the real part of the integrand was odd in $\xi_n$ and yielded zero on integration.
Identifying $-\mathrm{Im}\Sigma=\frac{1}{2\tau}\mathrm{sgn}(\omega)$, I arrive at
\begin{eqnarray}
\tau^{-1}&=&n_iU_0^2\frac{eB}{\pi}\sum_{n=0}^{N}\frac{1}{|v_{zn}|}\nonumber\\
&=&n_iU_0^2\frac{eB}{\pi}\sqrt{\frac{m}{2}}\sum_{n=0}^{N}\frac{1}{\sqrt{E_F-(n+1/2)\omega_c}}.
\label{sctime_supp}
\end{eqnarray}
which is quoted in the main text. The discrete sum is evaluated as before:
\begin{eqnarray}
\sum_{n=0}^{N}\frac{1}{\sqrt{E_F-(n+1/2)\omega_c}}&=&\sum_{n=0}^N\frac{1}{\sqrt{E_F-(n+1/2)\omega_c}}\nonumber\\
&=&\int_{\frac{1}{2}}^{X-\delta}\frac{1}{\sqrt{E_F-x\omega_c}}dx+\frac{1}{2}\left[\frac{1}{\sqrt{E_F-(X-\delta)\omega_c}}+\frac{1}{\sqrt{E_F-\omega_c/2}}\right]\nonumber\\
&+&\frac{1}{12}\left[\left(\frac{1}{\sqrt{E_F-x\omega_c}}\right)'\bigg\rvert_{x\rightarrow X-\delta}-\left(\frac{1}{\sqrt{E_F-x\omega_c}}\right)'\bigg\rvert_{x\rightarrow 1/2}\right]+\cdots\nonumber\\
&=&-\frac{2}{\omega_c}\left[(\delta\omega_c)^{1/2}-(E_F-\omega_c/2)^{1/2}\right]+\frac{1}{2}\left[\frac{1}{(\delta\omega_c)^{1/2}}+\frac{1}{(E_F-\omega_c/2)^{1/2}}\right]\nonumber\\
&+&\frac{\omega_c}{24}\left[\frac{1}{(\delta\omega_c)^{3/2}}-\frac{1}{(E_F-\omega_c/2)^{3/2}}\right]+\cdots.
\end{eqnarray}
Expanding in $\omega_c/E_F$,
\beq
\sum_{n=0}^{N}\frac{1}{\sqrt{E_F-(n+1/2)\omega_c}}\approx\left[2\frac{E_F^{1/2}}{\omega_c}\right]+\left[\frac{1}{48}\frac{\omega_c}{E_F^{3/2}}\right]+\left[\omega_c^{-1/2}\delta^{-3/2}\left(-2\delta^2+\frac{1}{2}\delta+\frac{1}{24}\right)\right].
\eeq
Inserting this in Eq.~(\ref{sctime_supp}), I finally have
\beq
\tau^{-1}\approx \frac{n_iU_0^2}{\pi}2^{1/2}m^{3/2}E_F^{1/2}\left[1+\frac{1}{96}\frac{\omega_c^2}{E_F^2}-\frac{\omega_c^{1/2}}{E_F^{1/2}}\frac{1}{\delta^{3/2}}\left(\delta^2-\frac{1}{4}\delta-\frac{1}{48}\right)\right].
\eeq
The first term is the zero field contribution, the second is the smooth contribution quadratic in field, and the third leads to oscillations. I discard the oscillating part and take the first two terms, which appear in the main text.

\begin{figure}
\includegraphics[angle=0,width=0.7\columnwidth,trim={0 6cm 0 0},clip]{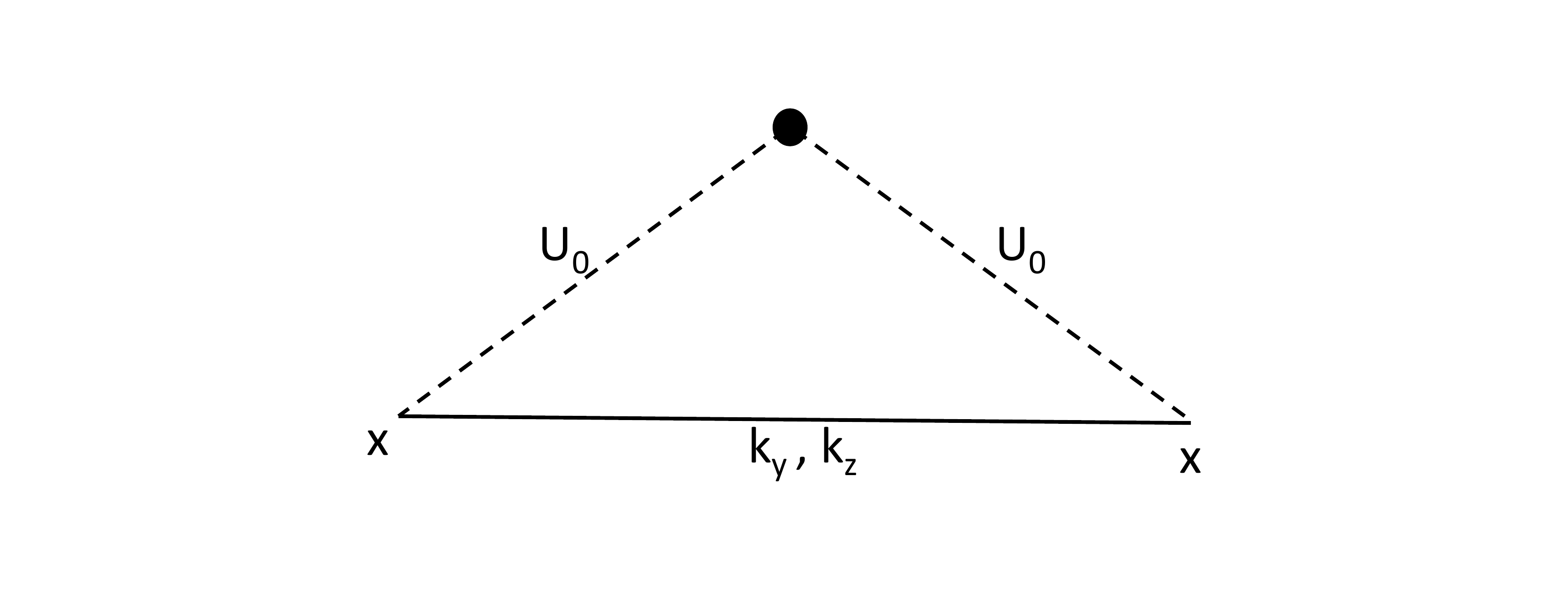}
\caption{Self-energy diagram to calculate $1/\tau$}
 \label{selfen_supp}
\end{figure}

\subsection{Calculation of $\sigma_{zz}$ for a general spectrum}

I now calculate $\sigma_{zz}(B)$ for an arbitrary choice of $\ve_{xy}$ and $\ve_z$. The Landau levels $\ve_{xy}(n)$ no longer have a simple analytical form. They are, instead, derived from the semiclassical quantization condition
\beq
Sl_B^2=2\pi(n+\gamma),
\eeq
where $S(\ve)$ is the area of the surface enclosed by the isoenergy contour $\ve_{xy}=\ve$ in the two-dimensional $k-$space, $l_B=\frac{1}{\sqrt{eB}}$ is the magnetic length, and $\gamma$ is the semiclassical phase. In the general case, when the dispersion is non-parabolic, two changes arise: $S$ is no longer the area of a circle and, more importantly, $\gamma$ is no longer a constant but a function of $\ve$ itself. In the simplest case where singularities in the isoenergy contours and interband effects can be ignored, it can be shown that
\beq
\gamma(\ve)-\frac{1}{2}=-\frac{\pi B}{2e}\frac{\partial \chi}{\partial\ve},\label{gamma_supp}
\eeq
where $\chi$ is the two-dimensional orbital magnetic susceptibility corresponding to $\ve_{xy}=\ve$. Here, it has been assumed that the band-bottom is parabolic so that $\gamma_0=\gamma(0)=1/2$. To compute the discrete sum in Eq.~(\ref{sigzz_supp}) with the help of Euler-Maclaurin formula as before, once again define $n+\gamma=x$ with $X$ as the value $x$ takes at $E_F$, and $\delta=X-(N+\gamma)$. The key point is, unlike in the parabolic case, now $dn=dx-(d\gamma/dx)dx$. This yield an extra term in the Euler-Maclaurin formula:
\begin{eqnarray}
\sum_{n=0}^N|v_{zn}|&=&\int_{\gamma_0}^{N+\gamma}|v_z(x)|dx-\int_{\gamma_0}^{N+\gamma}|v_z(x)|\frac{\partial\gamma(x)}{\partial x}dx+\frac{1}{2}[|v_z(N+\gamma)|+|v_z(\gamma_0)|]+\frac{1}{12}\left[\frac{\partial |v_z|}{\partial x}\bigg\rvert_{N+\gamma}-\frac{\partial |v_z|}{\partial x}\bigg\rvert_{\gamma_0}\right]+\cdots\nonumber\\
&=&\underbrace{\int_{\gamma_0}^{X-\delta}|v_z(x)|dx}_\text{T1}-\underbrace{\int_{\gamma_0}^{X-\delta}|v_z(x)|\frac{\partial\gamma(x)}{\partial x}dx}_\text{T2}+\underbrace{\frac{1}{2}[|v_z(X-\delta)|+|v_z(\gamma_0)|]}_\text{T3}+\underbrace{\frac{1}{12}\left[\frac{\partial |v_z|}{\partial x}\bigg\rvert_{X-\delta}-\frac{\partial |v_z|}{\partial x}\bigg\rvert_{\gamma_0}\right]}_\text{T4}+\cdots
\end{eqnarray}
Terms T1, T3, and T4 are identical to that appearing in the parabolic case, while term T2 is new. As before, only terms upto linear order in $B$ in the sum will be kept which is equivalent to keeping terms that contain $\frac{\partial |v_z|}{\partial x}$ and neglecting all higher order derivatives. This is true because $\frac{\partial |v_z|}{\partial x}=\frac{\partial |v_z|}{\partial \ve}\frac{\partial \ve}{\partial S}\frac{\partial S}{\partial x}=\omega_c(\ve)\frac{\partial |v_z|}{\partial \ve}$, where I have used $\frac{\partial S}{\partial \ve}=2\pi m(\ve)$ and $\frac{\partial S}{\partial x}=2\pi eB$. Expanding, one gets
\beq
\mathrm{T1}+\mathrm{T3}+\mathrm{T4}=\int_0^X| v_z(x)|dx+|v_z(E_F)|\left(\frac{1}{2}-\delta\right)+\frac{\omega_c(0)}{24}\frac{\partial |v_z|}{\partial \ve}\bigg\rvert_0+\frac{\omega_c(E_F)}{2}\frac{\partial |v_z|}{\partial \ve}\bigg\rvert_{E_F}\left(\delta^2-\delta+\frac{1}{6}\right).\label{t134}
\eeq
When $\ve=E_F$, Eq.~(\ref{dispersion_supp}) gives $\ve_z=0$, i.e., it is at its minimum. Assuming $\ve_z$ is an analytic function, it implies $v_z(E_F)=0$. Therefore, the second term above does not contribute. Note that the last term representing oscillations need not be well-behaved. The derivative $\frac{\partial |v_z|}{\partial \ve}\big\rvert_{E_F}$ may diverge, as happens for example in the parabolic case. This arises because the oscillating part need not be an analytic function of $\omega_c$ or $\delta$, as evidenced in Eq.~(\ref{parabolares_supp}). Since this is not the focus of the calculation and will be ignored anyway, I do not discuss it further. Next, I consider term T2. Integrating by parts and keeping terms upto linear order in $B$, I get
\beq
\mathrm{T2}=-|v_z(E_F)|\left(\gamma(E_F)-\frac{1}{2}\right)+\int_{0}^{E_F}\frac{\partial |v_z|}{\partial \ve}\left(\gamma(\ve)-\frac{1}{2}\right)d\ve .\label{t2}
\eeq
The first term is zero based on the arguments above. Collecting the terms from Eqs.~(\ref{t134}) and (\ref{t2}), I have
\beq
\sum_{n=0}^N|v_{zn}|=\left[\int_0^X |v_z(x)|dx\right]+\left[\underbrace{\frac{\omega_c(0)}{24}\frac{\partial |v_z|}{\partial \ve}\bigg\rvert_0}_\text{S1}+\underbrace{\int_{0}^{E_F}\frac{\partial |v_z|}{\partial \ve}\left(\gamma(\ve)-\frac{1}{2}\right)d\ve}_\text{S2}\right].\label{s1s2}
\eeq
Both terms S1 and S2 can be related to the two-dimensional orbital magnetic susceptibility $\chi$. For term S2 it is obvious: using Eq.~(\ref{gamma_supp}) S2 becomes $-\frac{\pi B}{2e}\int_{0}^{E_F}\frac{\partial |v_z|}{\partial \ve}\frac{\partial\chi}{\partial\ve}d\ve$. For S1, recall the thermodynamic definition, $\chi=-\frac{\partial ^2\Omega}{\partial B^2}$, where $\Omega$ is the grand potential defined at $T=0$ as $\Omega=\frac{eB}{2\pi}\sum_{n=0}^N[\ve_{xy}(n)-E_F]$. Comparing it with Eq.~(\ref{sigzz_supp}), it is obvious that the calculation of $\Omega$ is identical as $\sigma_{zz}(B)$ if one identifies $|v_{zn}|\rightarrow[E_F-\ve_{xy}(n)]$ so that $\frac{\partial |v_{z}|}{\partial \ve}\rightarrow -1$. Indeed, carrying out the sum for $\Omega$, one finds $\chi=-\frac{e^2}{12\pi m}$ \cite{sho}. Thus, S1 is simply $-\frac{\pi B}{2e}\frac{\partial |v_z|}{\partial \ve}\chi\bigg\rvert_0$, where I have used $\omega_c(0)=eB/m$. Rewriting S1 and S2 in terms of $\chi$ and plugging Eq.~(\ref{s1s2}) back into Eq.~(\ref{sigzz_supp}), the final expression is
\beq
\sigma_{zz}(B)\approx\sigma_{zz}(0)-\left[\frac{\partial |v_z|}{\partial\ve}\chi\bigg\rvert_{\ve=0}+\int_{0}^{E_F}\frac{\partial |v_z|}{\partial\ve}\frac{\partial\chi}{\partial\ve}d\ve\right]\frac{e^2\tau}{4\pi} B^2.
\eeq
This expression is quoted in the main text.

\subsection{Numerical calculation of Landau levels for a square lattice spectrum}

To calculate $\sigma_{zz}(B)$ numerically using Eq.~(\ref{sigzz_supp}), one needs to calculate the Landau level spectrum $\ve_{xy}(n)$ corresponding to $\ve_{xy}(k_x,k_y)$ in Eq.~(\ref{dispersion_supp}). In the main text, I considered the dispersion $\ve_{xy}(k_x,k_y)=4t-2t[\mathrm{cos}(k_xa)+\mathrm{cos}(k_ya)]$. The corresponding lattice Hamiltonian is that of a  square lattice with nearest neighbor interaction:
\beq
H=4t \sum_ic_i^{\dagger}c_i-\sum_{<i,j>}\left(t_{ij}c_i^{\dagger}c_j+h.c.\right),\label{tb_supp}
\eeq
with $t_{ij}=t$. Magnetic field is introduced via Peierls substitution for the hopping parameters as $t_{ij}=te^{ie\int_i^j \mb{A}.d\mb{l}}$, where $\mb{A}$ is the magnetic vector potential, and $d\mb{l}$ denotes an infinitesimal line element from points $i$ to $j$ on the lattice. I use the gauge $\mb{A}=(0,Bx,0)$. Writing $x$ as $la$, where $l$ is an integer, the phase in the hopping parameter becomes $e\int A_ydy=2\pi l\phi/\phi_0$, with $\phi$ being the magnetic flux and $\phi_0$ being the flux quantum. It is seen that for $\phi/\phi_0=p/q$, where $p$ and $q$ are integers, a periodicity of $qa$ in the $x$-direction is restored. In my calculations I take $p=1$. Going to the Fourier space, Eq.~(\ref{tb_supp}) can be cast in terms of a $q-$component basis $C=[c^1,\cdots,c^q]$ as
\beq
-tc^{n+1}_{k_x,k_y}e^{ik_xa}-tc^{n-1}_{k_x,k_y}e^{-ik_xa}-2tc^n_{k_x,k_y}[\mathrm{cos}(k_yb-2\pi n\phi)+4t], \quad n=1,\cdots, q.
\label{squarepart_supp}
\eeq
Thus, Eq.~(\ref{tb_supp}) becomes
\beq
H=\sum_{\mb{k}}
C^{\dagger}_{\mb{k}} \mathcal{H}_{\mb{k}}C_{\mb{k}}
\eeq
with $\mathcal{H}$ a $q\times q$ matrix given by (\ref{squarepart_supp}). The problem is thus reduced to an eigenvalue problem for a $q\times q$ matrix. Solving the eigenvalue problem numerically for  $k_x=k_y=0$, I get the discrete energy values for each value of $\phi=n/q$, $n=1,\cdots,q$ which gives us the Landau level spectrum $\ve_{xy}(n)$.

\end{widetext}


\begin{thebibliography}{99}


\bibitem{zim} J. M. Ziman, \emph{Electrons and Phonons: The Theory of Transport Phenomena in Solids} (Clarendon Press, Oxford, 1967).

\bibitem{pal} H. K. Pal and D. L. Maslov, Phys. Rev. B \tb{81}, 214438 (2010). 

\bibitem{pipbook} A. B. Pippard, \emph{Magnetoresistance in Metals} (Cambridge University Press, Cambridge, England, 1989).

\bibitem{pip} A. B. Pippard,Proc. R. Soc. London, Ser. A \tb{282}, 464 (1964).

\bibitem{nielsen} H. B. Nielsen and M. Ninomiya, Phys. Lett. B \tb{130}, 389 (1983).

\bibitem{burkov} A. A. Burkov, Phys. Rev. Lett. \tb{113}, 247203 (2014).

\bibitem{son} D. T. Son and B. Z. Spivak, Phys. Rev. B \tb{88}, 104412 (2013).

\bibitem{niu} Y. Gao, S. A. Yang, and Q. Niu, Phys. Rev. B \tb{95}, 165135 (2017).

\bibitem{andreev} A. V. Andreev and B. Z. Spivak, Phys. Rev. Lett. \tb{120}, 026601 (2018).

\bibitem{sondheimer} E. H. Sondheimer, Proc. R. Soc. London, Ser. A \tb{268}, 100 (1962).

\bibitem{goswami} P. Goswami, J. H. Pixley, and S. Das Sarma, Phys. Rev. B \tb{92}, 075205 (2015).

\bibitem{stroud} D. Stroud and F. P. Pan, Phys. Rev. B \tb{13}, 1434 (1976).

\bibitem{miller} D. L. Miller and B. Laikhtman, Phys. Rev. B \tb{54}, 10669 (1996).

\bibitem{parish} J. Hu, M. M. Parish, and T. F. Rosenbaum, Phys. Rev. B \tb{75}, 214203 (2007).

\bibitem{argyres} P. N. Argyres and E. N. Adams, Phys. Rev. \tb{104}, 900 (1956).

\bibitem{abrbook} A. A. Abrikosov, \emph{Fundamentals of the Theory of Metals} (Elsevier, Amsterdam, 1988).

\bibitem{fred} A. Raoux, F. Pi{\'e}chon, J.-N. Fuchs, and G. Montambaux, Phys. Rev. B \tb{91}, 085120 (2015).

\bibitem{abr} A. A. Abrikosov, JETP \tb{29}, 746 (1969).

\bibitem{comment} This implicitly assumes $E$ to be a monotonic function of $k_z$ which is not necessary. More generally, $\int dk_z\rightarrow \sum_l\frac{2}{|v_z^l|}\int d\xi_n$, where $l$ denotes all possible positive roots $E(n,k_z^l)=E_n$. In this case, $v_z$ appearing in Eqs.~(\ref{sigzz}) and (\ref{result}) should be replaced by $v_z^l$ and the expression should be summed over $l$.

\bibitem{roth} L. M. Roth, Phys. Rev. \tb{145}, 434 (1966).

\bibitem{sho} D. Shoenberg, \emph{Magnetic Oscillations in Metals}, Cambridge Univ. Press (1984).
 


\end{thebibliography}
\end{document}